\documentclass[11pt,oneside,english]{amsart}
\usepackage[T1]{fontenc}
\usepackage[latin9]{inputenc}
\usepackage{geometry}
\usepackage{amstext}
\usepackage{amsthm}
\usepackage{amssymb}

\makeatletter
\numberwithin{equation}{section}
\numberwithin{figure}{section}
\theoremstyle{plain}
\newtheorem{thm}{\protect\theoremname}[section]
\theoremstyle{remark}

\theoremstyle{plain}
\newtheorem{lem}[thm]{\protect\lemmaname}

\usepackage{hyperref}
\date{}

\usepackage[textsize=footnotesize,textwidth=0.85in]{todonotes}

\usepackage{tikz-cd}
\usepackage{amsmath, amssymb}

\makeatother

\usepackage{babel}
\providecommand{\lemmaname}{Lemma}
\providecommand{\remarkname}{Remark}
\providecommand{\theoremname}{Theorem}

\begin{document}
\title{Equivalence of quantizations of the dispersionless KdV hierarchy}
\author{Xavier Blot}
\begin{abstract}
Wang recently constructed a quantization of the dispersionless KdV hierarchy using the Heisenberg vertex algebra. Independently, in joint work with Rossi, we obtained a quantization of the dispersionless KdV hierarchy as the trivial Cohomological Field Theory case of the meromorphic differential hierarchies. In this note, we prove that the two constructions coincide.
\end{abstract}

\maketitle

\section{Introduction}

The equations of the dispersionless KdV hierarchy are
\begin{align*}
\frac{\partial u}{\partial t_{n}} & =\frac{u^{n}}{n!}u_{x},\qquad n\geq0.
\end{align*}
This hierarchy is Hamiltonian
\[
\frac{\partial u}{\partial t_{n}}=\left\{ u,\overline{h}_{n}\right\} ,
\]
where the Hamiltonians are given by
\[
\overline{h}_{n}=\int\frac{u^{n+1}}{\left(n+1\right)!}dx
\]
and the Poisson bracket, known as the first (or Gardner--Zakharov--Fadeev) Poisson bracket of KdV, is given on the space of local functionals by $\left\{ \overline{f},\overline{g}\right\} =\int\tfrac{\delta\overline{f}}{\delta u}\partial_{x}\tfrac{\delta\overline{g}}{\delta u}$. All notations are introduced below.

\medskip

A first quantization of the dispersionless KdV hierarchy with respect to the first Poisson bracket appeared independently in Pogrebkov's work \cite{pogrebkov2003boson,Pogrebkov_Hierarchy} and in Symplectic Field Theory \cite{eliashberg2010introduction,eliashberg2007symplectic}. A significant advance was achieved by Buryak and Rossi \cite{BR2016}, who provided a geometric interpretation of the quantum Hamiltonians in terms of intersection numbers with the double ramification cycle in the moduli space of curves. This construction is in fact more general, since it provides a quantization by deformation of an entire family of integrable hierarchies---one for each Cohomological Field Theory (CohFT). It led to several important developments, including the definition \cite{BDGR2} and study of quantum tau functions \cite{blot2022quantum,blot2024quantum,blot2024cohomological}, which, as a byproduct, contributed to the proof of the DR--DZ equivalence \cite{blot2024strong,blot2024master}, as key ideas emerged from the study of quantum tau functions. In another direction, it opened the way to the study of the spectral problem for these quantum Hamiltonians---an aspect already explored by Dubrovin \cite{dubrovin2016symplectic}---which has since revealed unexpected connections with modularity \cite{van2022quantum,van2024quantum}.

\medskip

Recently, two quantizations of the dispersionless KdV hierarchy have emerged that we now outline. 

\subsection{Meromorphic differentials hierarchies}

In a joint work with Rossi \cite{blot2024meromorphic}, we introduced a family of classical and quantum integrable hierarchies associated to arbitrary CohFTs, called the \emph{meromorphic differential hierarchies}. The Hamiltonian densities are defined as Hodge integrals integrated over  meromorphic differential strata. For the trivial CohFT, in the dispersionless limit, the \emph{meromorphic differential Hamiltonians} are
\begin{equation}
H_{d}^{\mathrm{MD}}(x)=\sum_{\substack{g,n\ge0\\
2g+n>0
}
}\frac{(-i\hbar)^{g}}{n!}\!\!\!\!\sum_{\substack{s_{1},\dots,s_{n}\ge0\\
s_{1}+\cdots+s_{n}=2g
}
}\!\!\!\!u_{s_{1}}\cdots u_{s_{n}}\,\mathrm{Coef}_{m_{1}^{\underline{s_{1}}}\cdots m_{n}^{\underline{s_{n}}}}\int_{\overline{\mathcal{H}}_{g}\!\left(-1,m_{1},\dots,m_{n},2g-1-\sum_{i=1}^{n}m_{i}\right)}\psi_{0}^{\,d+1},\label{eq:Hamilto-MD}
\end{equation}
for $d\geq-1$. Here:
\begin{itemize}
\item $\overline{\mathcal{H}}_{g}\left(m_{0},\dots,m_{n-1}\right)$ denotes the closure, in the moduli space $\overline{\mathcal{M}}_{g,n}$ of stable curves of genus $g$ with $n$ marked points (indexed from $0$ to $n-1$), of the locus of smooth marked curves $\left(C,x_{0},\dots,x_{n-1}\right)$ satisfying
\[
\omega_{C}\left(-\sum_{i=0}^{n-1}m_{i}x_{i}\right)\cong\mathcal{O}_{C},
\]
where $m_{0},\dots,m_{n-1}$ are integers satisfying $\sum_{i=0}^{n-1}m_{i}=2g-2$. If at least one $m_{i}$ is negative, $\overline{\mathcal{H}}_{g}\left(m_{0},\dots,m_{n-1}\right)$ is called a \emph{stratum of meromorphic differential}, and has codimension $g$. The class $\psi_{i}$, for $i=0,\dots,n-1$, is the first Chern class of the line bundle over $\overline{\mathcal{M}}_{g,n}$ whose fiber at a marked curve is the cotangent line at the $i$-th marked point. Note that no Hodge class appears as we are in the dispersionless limit. 
\item As shown in \cite[Proposition 5.1 and Theorem 1]{blot2024meromorphic}, the integral
\[
\int_{\overline{\mathcal{H}}_{g}\!\left(-1,m_{1},\dots,m_{n},\,2g-1-\sum_{i=1}^{n}m_{i}\right)}\psi_{0}^{\,d+1}
\]
is a polynomial of degree $2g$ in the variables $m_{1},\dots,m_{n}$. Moreover, once written in the falling factorial basis $m_{i}^{\underline{s_{i}}}=m_{i}\left(m_{i}-1\right)\cdots\left(m_{i}-s_{i}+1\right)$, it becomes homogeneous of degree $2g$. The notation $\mathrm{Coef}_{m_{1}^{\underline{s_{1}}}\cdots m_{n}^{\underline{s_{n}}}}$ means extracting the coefficient of $m_{1}^{\underline{s_{1}}}\cdots m_{n}^{\underline{s_{n}}}$ in this polynomial.
\item Finally, by a direct dimension argument, one verifies that the meromorphic differential Hamiltonians belong to the ring of \emph{differential polynomials} $\mathcal{A}\left[\hbar\right]:=\mathbb{C}\left[u_{0},u_{1},\dots;\hbar\right].$
\end{itemize}
Let $\partial_{x}:\mathcal{A}\left[\hbar\right]\rightarrow\mathcal{A}\left[\hbar\right]$ be the derivation $\partial_{x}=\sum_{i\geq0}u_{i+1}\tfrac{\partial}{\partial u_{i}}$. Denote by $\Lambda\left[\hbar\right]=\mathcal{A}\left[\hbar\right]/\left({\rm Im}\left(\partial_{x}\right)\oplus\mathbb{C}\right)$ the space of \emph{local functionals}. For a differential polynomial $f\in\mathcal{A}\left[\hbar\right]$, we denote its class in $\Lambda\left[\hbar\right]$ equivalently by $\overline{f}$ or $\int fdx$. 

In \cite{blot2024meromorphic}, we introduce a quantum bracket 
\[
\left[\cdot,\cdot\right]:\Lambda\left[\hbar\right]\times\Lambda\left[\hbar\right]\rightarrow\Lambda\left[\hbar\right]
\]
as a deformation quantization of the first Poisson bracket of the KdV hierarchy, with an explicit expression given in Proposition 2.6 therein. The integrability of the hierarchy is 
\[
\left[\overline{H_{d_{1}}^{{\rm MD}}},\overline{H_{d_{2}}^{{\rm MD}}}\right]=0,\qquad d_{1},d_{2}\geq-1,
\]
as shown in \cite[Proposition 3.5]{blot2024meromorphic}.

In the classical limit, the quantum Hamiltonians reduce to the classical ones: $\ensuremath{\overline{H_{d}^{\mathrm{MD}}}\big|_{\hbar=0}=\overline{h_{d}}}$, and the quantum bracket reduces to the first Poisson bracket of KdV: $\left\{ \overline{f},\overline{g}\right\} =\frac{1}{\hbar}\left.\left[\overline{f},\overline{g}\right]\right|_{\hbar=0}$.
\medskip

\subsection{Vertex algebra construction}

Shortly thereafter, Wang introduced in \cite{wang2024quantum} another quantization of the dispersionless KdV, this time using the Heisenberg vertex algebra. 

In this approach, he defines a quantum bracket deforming the first Poisson bracket of KdV; remarkably, it coincides with the quantum bracket of the meromorphic differential hierarchy (up to a rescaling $\hbar\rightarrow-i\hbar$ and an overall factor of $i$), see \cite[Theorem 1.1]{wang2024quantum}. 

Furthermore, Wang constructs a quantum deformation of the dispersionless KdV Hamiltonians densities, denoted by $H_{d}^{{\rm Wang}}$ for $d\geq-1$. He establishes recursive formulas for their computation and provides two explicit expressions for the Hamiltonian densities. One of them, proved in \cite[Theorem 3.4]{wang2024quantum}, expresses these densities as
\begin{equation}
H_{d}^{{\rm Wang}}=\left.\sum_{k=0}^{d+1}\frac{\left(-1\right)^{d+1-k}}{\left(d-k+2\right)!}\partial_{x}^{d+1-k}S_{\left(k+1\right)}\right|_{u_{j}\rightarrow(\left(\sqrt{-i\hbar}\right)^{j}u_{j}},\label{eq:Hamilto-Wang}
\end{equation}
where $\ensuremath{S_{(k+1)}=[z^{k+1}]\,\exp\!\big(\sum_{j\ge0}\tfrac{u_{j}}{(j+1)!}\,z^{j+1}\big)}$. The notation $u_{j}\rightarrow(\sqrt{-i\hbar})^{j}u_{j}$ indicates the substitution of each formal variable $u_{j}$ by $(\sqrt{-i\hbar})^{j}u_{j}$. The appearance of $\sqrt{-i\hbar}$ poses no issue, since only even powers occur in $H_{d}^{{\rm Wang}}$. Notice that we rescaled Wang's convention by $\hbar\rightarrow-i\hbar$. Wang proves that the Hamiltonians mutually commute with respect to the quantum bracket:
\[
\left[\overline{H_{d_{1}}^{{\rm Wang}}},\overline{H_{d_{2}}^{{\rm {\rm Wang}}}}\right]=0,\qquad d_{1},d_{2}\geq-1,
\]
where the bracket $\left[\cdot,\cdot\right]$ uses the convention of the meromorphic differential hierarchy. 

In the classical limit, the quantum Hamiltonians reduce to the classical ones: $H_{d}^{{\rm Wang}}\big|_{\hbar=0}=h_{d}$.

\medskip

\subsection{Identification}

In this paper, we establish the following identification. 
\begin{thm}
\label{thm:main}For all $d\geq-1$, we have
\[
H_{d}^{{\rm MD}}=H_{d}^{{\rm Wang}}.
\]
\end{thm}

Since the quantum brackets coincide, the two corresponding quantum hierarchies are identical.

\medskip

\subsection{Comparison with other quantizations of dispersionless KdV}

In \cite{blot2024meromorphic}, we constructed another family of classical and quantum integrable hierarchies, called the \emph{twisted double ramification hierarchies}. The quantum bracket coincides with the one of the meromorphic differential hierarchy, while the Hamiltonian densities are also given as Hodge integrals, this time integrated over the twisted double ramification cycle. For the trivial CohFT, in the dispersionless limit, the two integrals coincide; consequently the Hamiltonian densities of the twisted DR hierarchy satisfy 
\[
H_{d}^{{\rm DR}^{1}}=H_{d}^{{\rm MD}},
\]

as shown in \cite[Propositon 5.3]{blot2024meromorphic}. 

Another quantization is provided by the quantum (untwisted) double ramification hierarchy \cite{BR2016} for the trivial CohFT in the dispersionless limit. The relation with the quantization given by the meromorphic differential hierarchy is exposed in \cite{blot2024meromorphic} (and more generally in the dispersive case, and even for the Hodge CohFT in \cite{blot2025meromorphic-hodge}). The quantum brackets are compared in \cite[Remark 2.7]{blot2024meromorphic} and the quantum Hamiltonians are related by
\[
H_{d}^{{\rm MD}}=\left(H_{d}^{{\rm DR}}\right)^{\left[0\right]},
\]
as shown in \cite[Section 5.3]{blot2024meromorphic}, where the upper index $[0]$ stands for extracting the degree $0$ of the Hamiltonian densities of the double ramification hierarchy, with respect to the grading $\deg u_{i}=i,$ $\deg\hbar=-2.$ 

To summarize, we have
\begin{center}
\begin{tikzcd}[column sep=normal, row sep=normal]
{H_d^{\mathrm{DR}^1}}
  \arrow[r, equal, shorten <=20pt, shorten >=20pt] &
{H_d^{\mathrm{MD}}}
  \arrow[r, equal, shorten <=20pt, shorten >=20pt] &
{(H_d^{\mathrm{DR}})^{[0]}} \\
& {H_d^{\mathrm{Wang}}}
  \arrow[u, equal, "\textrm{ this paper}"', shorten <=5pt, shorten >=5pt]
\end{tikzcd}
\end{center}

\medskip

%

\subsection{Further developments}
This note identifies a link between intersection-theoretic and vertex-algebraic approaches to a quantization of the dispersionless KdV hierarchy. Further aspects of this connection are expected to emerge and may be included in an extended version of this work.

\medskip

\subsection{Acknowledgments}

The author was supported by the grant OCENW.M.21.233 of the Dutch Research Council.

\bigskip

\section{Proof of Theorem~\ref{thm:main}}

We first establish the identity for local functionals
\[
\overline{H_{d}^{{\rm MD}}}=\overline{H_{d}^{{\rm Wang}}},\qquad d\geq-1.
\]
We endow the polynomial differentials and local functionals with the grading 
\[
\deg u_{i}=i,\quad\deg\hbar=-2.
\]
The equality follows from the following lemma. 
\begin{lem}
[Lemma 8.1  in \cite{blot2024meromorphic}]\label{lem:u3}Let 
\[
\overline{H}=\int\left(\frac{u^{3}}{3!}+O\left(\hbar\right)\right)dx,\qquad\overline{Q}=\int\left(Q_{0}+O\left(\hbar\right)\right)dx
\]
be two local functionals of homogeneous degree $0$ such that $\left[\overline{Q},\overline{H}\right]=0.$ Then $\overline{Q}$ is uniquely determined by $\overline{H}$ and $\overline{Q_{0}}$. In particular, if $\overline{Q_{0}}=0$ then $\overline{Q}=0$. 
\end{lem}

We now observe the following: 
\begin{itemize}
\item The first Hamiltonian coincides in both constructions:
\[
\overline{H_{1}^{{\rm MD}}}=\int\frac{u^{3}}{6}dx=\overline{H_{1}^{{\rm Wang}}}.
\]
For the meromorphic differential hierarchy, this follows from a dimension count and from the fact that, in genus $0$, the stratum of meromorphic differentials coincides with the entire moduli space $\overline{\mathcal{M}}_{0.5}$, for which $\int_{\overline{\mathcal{M}}_{0,5}}\psi_{0}^{2}=1$. For Wang\textquoteright s hierarchy, the same expression follows directly from the definition Eq.~(\ref{eq:Hamilto-Wang}). 
\item Each Hamiltonian has homogeneous degree $0$. For the Hamiltonians of the meromorphic differential hierarchy, this follows directly from their definition Eq.~(\ref{eq:Hamilto-MD}), while for those of Wang\textquoteright s hierarchy it follows from the homogeneity properties established in \cite[Lemma 3.2]{wang2024quantum}. Moreover, in the classical limit $\left(\hbar=0\right)$, the Hamiltonian densities coincide:
\[
\left.H_{d}^{{\rm MD}}\right|_{\hbar=0}=\left.H_{d}^{{\rm Wang}}\right|_{\hbar=0}=\frac{u^{d+1}}{\left(d+1\right)!}.
\]
\item Both hierarchies are endowed with the same quantum bracket and they are both integrable.
\end{itemize}
Combining these observations with Lemma~\ref{lem:u3}, we conclude that $\overline{H_{d}^{{\rm MD}}}=\overline{H_{d}^{{\rm Wang}}}$ for $d\geq-1$. 

We then deduce that $H_{d}^{{\rm MD}}=H_{d}^{{\rm Wang}}$ for $d\geq-1$, using the identities
\[
\frac{\delta\overline{H_{d}^{{\rm MD}}}}{\delta u}=H_{d-1}^{{\rm MD}},\qquad\frac{\delta\overline{H_{d}^{{\rm Wang}}}}{\delta u}=\overline{H_{d-1}^{{\rm Wang}}},
\]
valid for all $d\geq0$. The first identity is proved in a slightly more general context in \cite[Section 8.1]{blot2024meromorphic}, while the second is established below.

Since a derivative with respect to $\partial_{x}$ lies in the kernel of the integration map, we obtain using Eq.~(\ref{eq:Hamilto-Wang}),
\[
\frac{\delta\overline{H_{d}^{{\rm Wang}}}}{\delta u}=\frac{\delta}{\delta u}\int\left.S_{\left(d+1\right)}\right|_{u_{j}\rightarrow\left(\sqrt{-i\hbar}\right)^{j}u_{j}}dx.
\]
Hence
\[
\frac{\delta\overline{H_{d}^{{\rm Wang}}}}{\delta u}=\sum_{s\geq0}\left(-\partial_{x}\right)^{s}\frac{\partial}{\partial u_{s}}\left(\left.S_{\left(d+1\right)}\right|_{u_{j}\rightarrow\left(\sqrt{-i\hbar}\right)^{j}u_{j}}\right).
\]
A direct computation shows that 
\[
\frac{\partial}{\partial u_{s}}\left.S_{\left(d+1\right)}\right|_{u_{j}\rightarrow\left(\sqrt{-i\hbar}\right)^{j}u_{j}}=\begin{cases}
\frac{\sqrt{-i\hbar}^{s}}{\left(s+1\right)!}\left.S_{\left(d-s\right)}\right|_{u_{j}\rightarrow\left(\sqrt{-i\hbar}\right)^{j}u_{j}} & s=0,\dots,d\\
0 & \textrm{otherwise. }
\end{cases}
\]
This completes the proof.

\bibliographystyle{alpha}
\bibliography{qKdV}

\end{document}